\documentclass{elsart}
\usepackage{psfig}
\def\deg{\ifmmode^{\circ}\;\else$^{\circ}\;$\fi} % overwrites \deg in LaTeX
\def\lsim{\,\lower2truept\hbox{${<
 \atop\hbox{\raise4truept\hbox{$\sim$}}}$}\,}
\def\gsim{\,\lower2truept\hbox{${>
 \atop\hbox{\raise4truept\hbox{$\sim$}}}$}\,}

\begin{document}

\runauthor{De Zotti, Gruppioni, Ciliegi, Burigana and Danese}
\begin{frontmatter}
\title{Polarization fluctuations due to extragalactic sources}

\author[Padova]{G.~De~Zotti\thanksref{offprints}} 
\author[Bologna]{C.~Gruppioni} 
\author[Bologna]{P.~Ciliegi} 
\author[tesre]{C.~Burigana}
\author[sissa]{L.~Danese}
\thanks[offprints]{{\it Send offprint requests to:} dezotti@pd.astro.it}

\address[Padova]{Osservatorio Astronomico di Padova, Vicolo dell'Osservatorio 
5, I-35122 Padova, Italy}
\address[Bologna]{Osservatorio Astronomico di Bologna, Via Ranzani 1, I-40127
Bologna, Italy}
\address[tesre]{Istituto TESRE/CNR, Via Gobetti 101, I-40129 Bologna, Italy}
\address[sissa]{SISSA, International School for Advanced Studies, Via Beirut
2-4, I-34014 Trieste, Italy}

\begin{abstract}
We have derived the relationship between polarization and intensity 
fluctuations due to point sources. In the case of a 
Poisson distribution of a population with uniform evolution properties
and constant polarization degree, polarization fluctuations are simply equal
to intensity fluctuations times the average polarization degree. Conservative 
estimates of the polarization degree of the classes of extragalactic sources 
contributing to fluctuations in the frequency ranges covered by the
forthcoming space missions MAP and Planck Surveyor indicate that 
extragalactic sources will not be a strong limiting factor to 
measurements of the polarization of the Cosmic Microwave Background.
\end{abstract}
\end{frontmatter}

\section{Introduction}
There are good prospects that the forthcoming space missions designed to 
provide high sensitivity and high resolution maps of the cosmic 
microwave background (CMB) will also measure the CMB polarization 
fluctuations (Knox 1998; Bouchet et al. 1999).

The current design of instruments for the Planck Surveyor mission 
(the third Medium-sized mission of ESA's Horizon 2000 Scientific
Programme)  
provides good sensitivity to polarization at all LFI (Low Frequency 
Instrument) frequencies (30, 44, 70, and 100 GHz) as well as at three
HFI 
(High Frequency Instrument) frequencies (143, 217 and 545 GHz).
The NASA's MIDEX class mission MAP has also polarization sensitivity 
in all channels (30, 40 and 90 GHz).

The extraction of the very weak cosmological polarization signal 
requires both a great sensitivity of the instruments and a careful
control of foregrounds. An analysis of the effect of Galactic 
polarized emissions (synchrotron and dust) on CMB 
measurements by the Planck and MAP missions was carried out by Bouchet 
et al. (1999). So far, however, the effect of extragalactic sources was not 
considered. On the other hand, significant linear 
polarization is seen in most compact, flat-spectrum radio sources which 
are the main contributors to small scale foreground intensity
fluctuations 
at $\lambda > 1\,$mm (Toffolatti et al. 1998) and the thermal dust
emission 
from galaxies which dominate the counts at sub-mm wavelengths is also 
expected to be polarized to some extent, as the Galactic dust emission
is 
observed to be (Hildebrand 1996).

In this paper we derive the relationship between intensity and 
polarization fluctuations in the case of a Poisson distribution of 
point sources, discuss the polarization degree of the 
relevant classes of extragalactic sources, and exploit recent 
evolutionary models to estimate the power spectrum of polarization 
fluctuations produced by them.

\section{Polarization fluctuations from a Poisson distribution of
polarized 
point sources}
Following Burn (1966) we define the complex linear polarization of a
source 
as $P_s= \Pi\exp(2i\chi)$ where $\Pi$ and $\chi$ are the degree and the 
angle of polarization, respectively. If the polarization angles of
different 
sources within any given solid angle element $d\Omega$ are uncorrelated, 
the expected value $\langle P_s \rangle$ is 0 and the variance is:
\begin{equation}
\sigma^2_P = {1/\pi}\int^{\pi}_{0} d\chi\, (P_s-\langle P_s \rangle)^2 = 
{1/\pi}\int^{\pi}_{0} d\chi\, \Pi^2\left[\cos^2(2\chi) + 
\sin^2(2\chi)\right] = \Pi^2.
\end{equation}
Let $N = n(S)\,dS\,d\Omega$ be the number of sources with flux 
$S$ within $dS$ and polarization degree $\Pi$ in a given solid angle
element 
$d\Omega$. As far as the central limit theorem holds, the expected value 
of the linear polarization within $d\Omega$ is also 0, with variance  
\begin{equation}
\sigma^2_{P,{d\Omega}} = {\Pi^2 \over {N}}. 
\end{equation}
The fluctuation amplitude of the polarized flux $S_P = {N} S P_s$ 
among the different cells of the sky subtending a 
solid angle $d\Omega$, due to sources with flux $S$, is therefore
obtained 
integrating over the probability distribution of $N$, $\rho(N)$: 
\begin{equation}
\sigma^2_{S_P,d\Omega}  =  \langle \left(S_P - \langle S_P
\rangle\right)^2  
\rangle = \langle S_P^2 \rangle  = 
 \int_0^\infty d{N}\, \rho({N}) {N} S^2 \Pi^2 = 
\Pi^2 S^2 \langle {N(S)} \rangle.
\end{equation}
In the case of a Poisson distribution of sources the variance is equal 
to the mean. Therefore, 
integrating the above equation over $S$ and over the solid 
angle we straightforwardly obtain $\sigma_{I_P} = \Pi \sigma_{I}$, where 
$\sigma_{I}$ is the rms intensity fluctuation for a Poisson distribution
of 
sources. 

The assumption of an equal polarization degree for all sources is
obviously 
unrealistic. However, it follows from the above calculations that, 
if the polarization degree is uncorrelated with flux, the result depends 
only on the mean value of $\Pi$. 

A dependence of the mean value of $\Pi$ on $S$ arises, in particular, 
in the case of contributions from 
classes of sources with different polarization properties and 
different shapes of the $\log N$--$\log S$ curves.  
These different populations must be dealt with separately.

\section{Linear polarization properties of the relevant classes of
sources}
Studies of the polarization properties of extragalactic sources at mm 
wavelengths are still scanty. Table 1 summarizes the main results 
on radio/mm sources. 
For each of the main source classes (in column 1) and for each
wavelength
in column 2, we give the median (column 3), the minimum (column 4) and
the
maximum (column 5) polarization value found in literature. Column 6
lists the references from which the reported values are derived.

\begin{table}
\centering
\caption{Radio and mm polarization measurements for extragalactic
sources}
\begin{tabular}{lrrrrl}
 & & & \\ \hline \hline
% & & & \\
 & \multicolumn{1}{c}{Band} & \multicolumn{1}{c}{P$_{\rm med}$}  &
 \multicolumn{1}{c}{P$_{\rm min}$ } & \multicolumn{1}{c}{P$_{\rm max}$}
& {Reference} \\
 &  & \multicolumn{1}{c}{(\%)}  & \multicolumn{1}{c}{(\%)} & 
\multicolumn{1}{c}{(\%)}
&  \\  \hline
% & & & \\
 BL Lac     & 4.8 GHz        &  3.6    &  1.5  &  7.5 & Aller et al.
1999 \\
            & 8.44 GHz       &  1.5    &  0.4  &  2.7 & Marcha et al.
1996 \\
            & 14.5 GHz       &  5.0    &  1.1  & 11.4 & Aller et al.
1999 \\
            & $0.8-1.1$ mm   &  7.1    &  3.0  & 12.8 & Nartallo et al.
1998 \\
            & $0.8-1.1$ mm   & 10.8    &  3.0  & 17.0 & Stevens et al.
1996 \\
\hline
 FS QSO     & $1.4-90$ GHz     &  2.5    &       &      & Saikia \& Salter
1988\\
 HPQ        & $0.8-1.1$ mm   &  7.4    &  4.6  & 10.5 & Nartallo et al.
1998\\
 LPQ        & $0.8-1.1$ mm   &  4.9    &  2.1  &  7.9 & Nartallo et al.
1998 \\
\hline
 FRII       & 1.4 GHz        &$\sim$4  &       &      & Saikia \& Salter
1988\\
            & 1.4 GHz        &$\sim$10 &  2.8  & 18.4 & Ishwara-Chandra
et al. 
1998 \\
            & 5 GHz          &$\sim$6  &       &      & Saikia \& Salter
1988\\
            & 5 GHz          &$\sim$10 &  2.4  & 18.2 & Ishwara-Chandra
et al. 
1998 \\
 FRI        & 5 GHz          &$\sim$7.5&       &      & Saikia \& Salter
1988\\
\hline \hline
\end{tabular}

\end{table}

As shown in Table 1, the three classes of objects (BL Lacs, QSOs and
bright radio galaxies) emit almost the same fraction of polarized
radiation in the radio band. 
The major extragalactic contributors to the non-thermal polarization  
at sub--mm and mm wavelengths are the flat spectrum (spectral
index 
$\alpha \leq 0.5$ if $S_{\nu} \propto \nu^{-\alpha}$) compact radio
sources
(mainly BL Lacertae objects and flat spectrum QSOs). These objects
constitute
about 50\% of the flux limited ($S > 1$ Jy at 5 GHz) radio catalogue
compiled by K\"uhr et al. (1981). Stickel et al. (1991) have drawn from this 
catalogue a complete sample of BL Lacertae objects brighter than 
$m = 20$ mag. Out of it, we have selected a complete 
sub--sample of 14 objects ($RA > 9^h$ and $\delta > 0^{\circ}$), for which
radio and/or mm polarization measurements are available. 

The polarization data for objects in this sub-sample 
(hereinafter {\it Stickel north}) are listed in Table 2. The tabulated 
percentage polarization degrees were obtained 
averaging the measurements from long term monitoring programs.
Data at 4.8 GHz, 8.1 GHz and 14.5 GHz are from Aller et al. (1985, 1999), 
those at 22 GHz, 31 GHz and 90 GHz are from Rudnick et al. (1985) and 
those at 1.1 mm are from Nartallo et al. (1998). 

\begin{table}
\parbox{12cm}{
\caption{Fractional Polarization Measurements for the {\it Stickel north}  
BL Lac Sample at several frequencies (GHz)}
}
\begin{tabular}{lcccccccr}
 & & & \\ \hline \hline
 & & & \\
\multicolumn{1}{c}{name} & \multicolumn{1}{c}{P(4.8)} &
\multicolumn{1}{c}{P(8.1)} & \multicolumn{1}{c}{P(14.5)} &
\multicolumn{1}{c}{P(22)} & \multicolumn{1}{c}{P(31)} & 
\multicolumn{1}{c}{P(90)} & \multicolumn{1}{c}{P(272)} &
\multicolumn{1}{c}{z} \\
 & \multicolumn{1}{c}{(\%)} & \multicolumn{1}{c}{(\%)} 
& \multicolumn{1}{c}{(\%)} &
\multicolumn{1}{c}{(\%)} & \multicolumn{1}{c}{(\%)} 
& \multicolumn{1}{c}{(\%)} &
\multicolumn{1}{c}{(\%)} & \\
 & & & \\ \hline
 & & & \\
0954+658 & 7.5 & $-$ & 6.0 & $-$ & $-$ & $-$ & $-$ & 0.367 \\
1147+245 & 2.4 & 3.3 & 4.6 & $-$ & $-$ & $-$ & $-$ & --    \\
1308+326 & 2.3 & 2.5 & 3.0 & 5.6 & 5.3 & 2.0 & 9.9 & 0.997 \\
1418+546 & 1.9 & 2.4 & 2.6 & $-$ & $-$ & 5.3 & $-$ & 0.152 \\
1538+149 & 4.9 & 5.8 & 7.5 & $-$ & $-$ & $-$ & $-$ & 0.605 \\
1652+398 & 1.7 & 3.1 & 3.0 & $-$ & $-$ & $-$ & $-$ & 0.033 \\
1749+096 & 3.1 & 3.3 & 3.4 & 6.5 & $-$ & 5.5 & 5.6 & 0.320 \\
1749+701 & 3.6 & 5.6 & 8.1 & $-$ & $-$ & $-$ & $-$ & 0.770 \\
1803+784 & 3.1 & 3.0 & 3.9 & $-$ & $-$ & $-$ & $-$ & 0.684 \\
1807+698 & 2.1 & 4.1 & 2.5 & $-$ & $-$ & $-$ & $-$ & 0.051 \\
1823+568 & 4.3 & 5.9 & 5.8 & $-$ & $-$ & $-$ & $-$ & 0.664 \\
2007+777 & 3.0 & 5.1 & 7.1 & $-$ & $-$ & $-$ & $-$ & 0.342 \\
2200+420 & 5.0 & 3.6 & 5.2 & 0.8 & 3.2 & 4.3 & 8.1 & 0.069 \\
2254+074 & 6.3 & $-$ &11.4 & $-$ & $-$ & $-$ & $-$ & 0.190 \\
\hline \hline
\end{tabular}
\end{table}

We have compared the average polarization percentages at the various 
wavelengths of sources in the {\it Stickel north} sample and  
in the sample by Aller et al. (1999), excluding those in 
common with the {\it Stickel north} sample.  The agreement is very good 
(see Fig. 1). The mean polarization percentages (P$_{m}$), 
the 68\% confidence uncertainties ($\Delta P_m$) obtained from the Student's $t$ 
distribution ($\Delta P_m= t_{\nu}(0.32) \sigma_P\,(N-1)^{-1/2}$, where $N$ 
is the number of sources in the bin and $\nu = N-1$) and the dispersions, 
$\sigma_{P}$, for the {\it Stickel north} sample are given in Table 3.

\begin{table}
\centering
\caption{Average Polarization Percentages for {\it Stickel north} Sample}
\begin{tabular}{cccccc}
 & &  \\ \hline \hline
 & &  \\
    $\lambda$ & $\nu$  &  P$_{m}$  & $\Delta P_m$  & $\sigma_P$ & N  \\
     (cm)     & (GHz)  &  (\%)     &  (\%)       &      (\%)  &    \\
              &        &           &             &            &    \\ \hline
              &        &           &             &            &    \\ 
      6.25    & 4.80   &   3.66    &   0.49      &     1.74   & 14 \\
      3.70    & 8.10   &   3.98    &   0.40      &     1.29   & 12 \\
      2.07    & 14.5   &   5.29    &   0.73      &     2.57   & 14 \\
      1.36    & 22.0   &   4.30    &   2.83      &     3.06   & ~3 \\
      0.97    & 31.0   &   4.25    &   2.69      &     1.48   & ~2 \\
      0.33    & 90.0   &   4.28    &   1.10      &     1.60   & ~4 \\
      0.11    & 272~   &   7.87    &   2.00      &     2.16   & ~3 \\ 
\hline \hline
\end{tabular}
\end{table}

Since for most objects in the two samples 
redshift information is available, we 
have converted the observed into the rest--frame wavelengths and 
computed the average polarization degrees in wavelength bins.
The results for the {\it Stickel north} sample are given in Table 4
and plotted in Figure 1$b$, where the results for the Aller sub-sample are 
also reported for comparison. Table 4 gives the 
adopted wavelength intervals, the average $\lambda$ in each interval, 
computed as the geometric mean of the two wavelength limits, the average 
polarization fraction with its 68\% confidence uncertainty from the 
Student's $t$ distribution and its dispersion, and the number of available 
measurements in each wavelength interval. 

\begin{figure}
\centerline{\psfig{figure=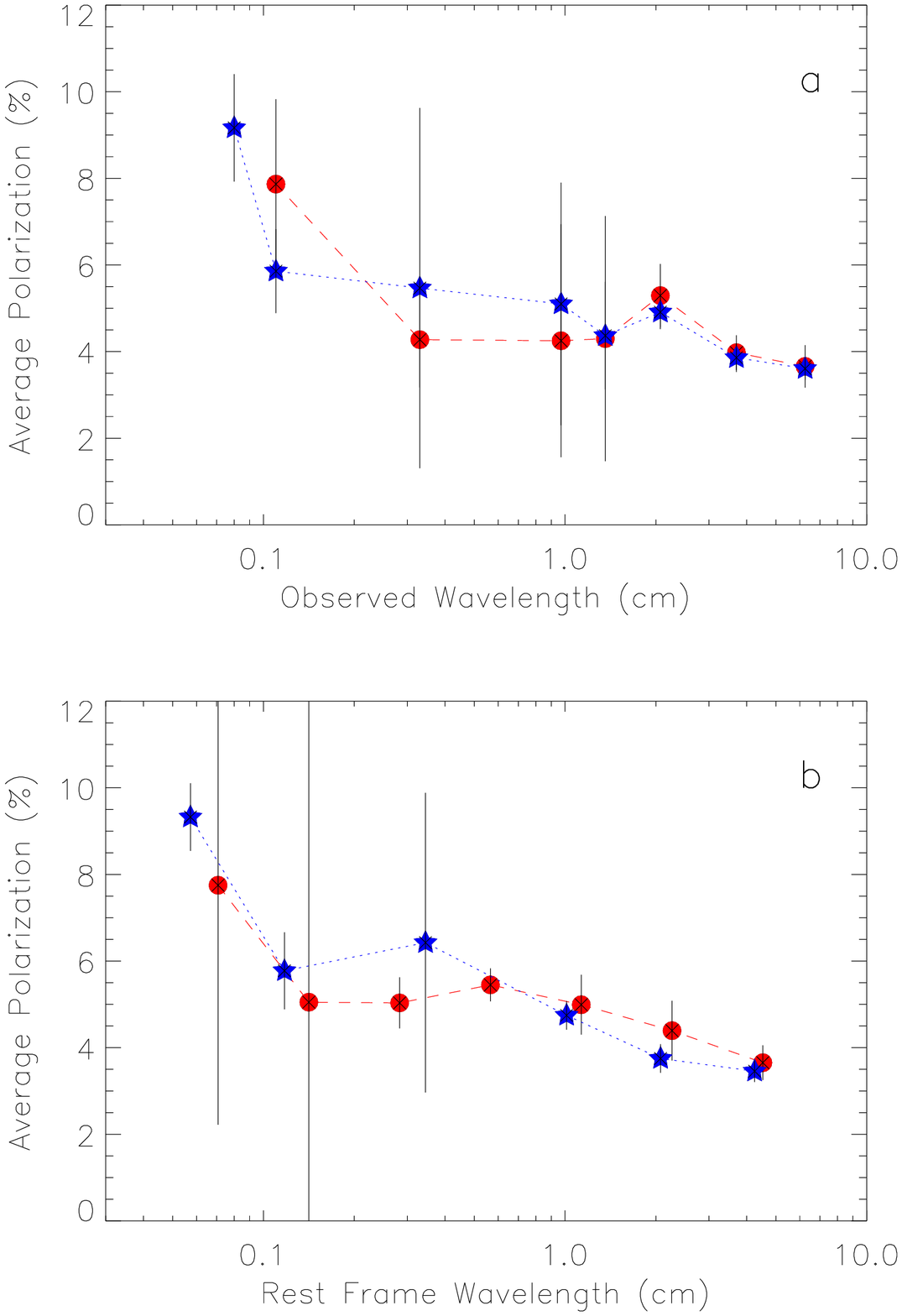,width=14cm} }
\caption{\label{fig1} Average fractional polarization as a function of 
wavelength for two samples of BL Lac objects: red filled circles 
connected by the red dashed line stand for the {\it Stickel north} sample,
while the blue filled stars connected by the blue dotted line stand for the
the 28 sources in the Aller et al. (1999) 
sample that are not included in the {\it Stickel north} sample.
The error bars correspond to the 68\% confidence uncertainties 
on the estimated mean polarization.} 
\end{figure}

\begin{table}
\centering
\caption{Average rest-frame polarization percentage for the 
{\it Stickel north} sample}
\begin{tabular}{cccccc}
 & &  \\ \hline \hline
 & &  \\
 $\Delta \lambda$ & $\lambda_{m}$ & P$_m$ & $\Delta P_m$ & $\sigma_P$ & N \\
 (cm)             & (cm)          & (\%)  & (\%)    & (\%)      &   \\
& &  \\ \hline
& &  \\
3.20--6.40 &  4.52   &    3.65     & 0.40  & 1.57   & 17  \\
1.60--3.20 &  2.26   &    4.39     & 0.69  & 2.43   & 14  \\
0.80--1.60 &  1.13   &    4.99     & 0.69  & 2.19   & 12  \\
0.40--0.80 &  0.57   &    5.45     & 0.38  & 0.21   &  2  \\
0.20--0.40 &  0.28   &    5.03     & 0.59  & 0.64   &  3  \\
0.10--0.20 &  0.14   &    5.05     & 7.83  & 4.31   &  2  \\
0.05--0.10 &  0.07   &    7.75     & 5.53  & 3.04   &  2  \\
\hline \hline
\end{tabular}
\end{table}

Measurements of polarized thermal emission from dust 
are only available for interstellar clouds in our own Galaxy. 
The distribution of observed polarization degrees of dense clouds 
at 100$\,\mu$m shows a peak at $\sim 2\%$ (Hildebrand 1996). 
The polarization degree of emission of silicate grains 
in clouds opaque to visible light but optically thin in the far-IR 
is nearly independent of wavelength if $\lambda \gg a$, $a$ being 
the grain size (Hildebrand 1988). This condition is very likely to be  
met at the long wavelengths of interest here. 

Polarization maps of the Orion molecular cloud at $100\,\mu$m, 
$350\,\mu$m, $450\,\mu$m, $1.3\,$mm, and $3.3\,$mm (Schleuning 1998; 
Rao et al. 1998) look very similar except at clumps of higher optical 
depth, where the polarization increases with wavelength.   
In fact, the maximum polarization decreases rapidly with increasing  
optical depth (Hildebrand 1996); thus, the mm/sub-mm polarization may 
be higher than at $100\,\mu$m. On the other hand, the overall
polarization 
of the light from a galaxy is the average of contributions from regions 
with different polarizing efficiencies and different orientations of 
the magnetic field with respect to the plane of the sky; all this works
to decrease the polarization level in comparison with the mean of
individual clouds. 

We could not find any published polarization measurement of dust
emission in 
external galaxies not hosting strong nuclear activity. 
However, the first results of the SCUBA Polarimeter include imaging 
polarimetry of the starburst galaxy M$\,$82 at 850$\,\mu$m with a 15'' 
beam size; a polarization degree of about 2\% was measured (J. Greaves, 
private communication). An early image, available at the 
polarimeter Web page
(http://www.jach.hawaii.edu/JCMT/scuba/scupol/\\
general/m82{\_}polmap.gif), 
shows very ordered polarization vectors over scales of a few hundred
parsecs. 
This may not occur in general. We may expect that, in other galaxies, 
the polarization vectors from giant molecular clouds are 
randomly ordered and  tend to cancel out. 

Polarimetric measurements at $170\,\mu$m with ISOPHOT have been carried
out
for two galaxies (U. Klaas, private communication): NGC$\,$1808 (P.I.:
E. 
Kr\"ugel) and NGC$\,$6946 (P.I.: G. Bower). The results, however, are
not 
available yet.

\section{Power spectrum of polarization fluctuations due to
extragalactic sources}

\begin{figure}
\centerline{\psfig{figure=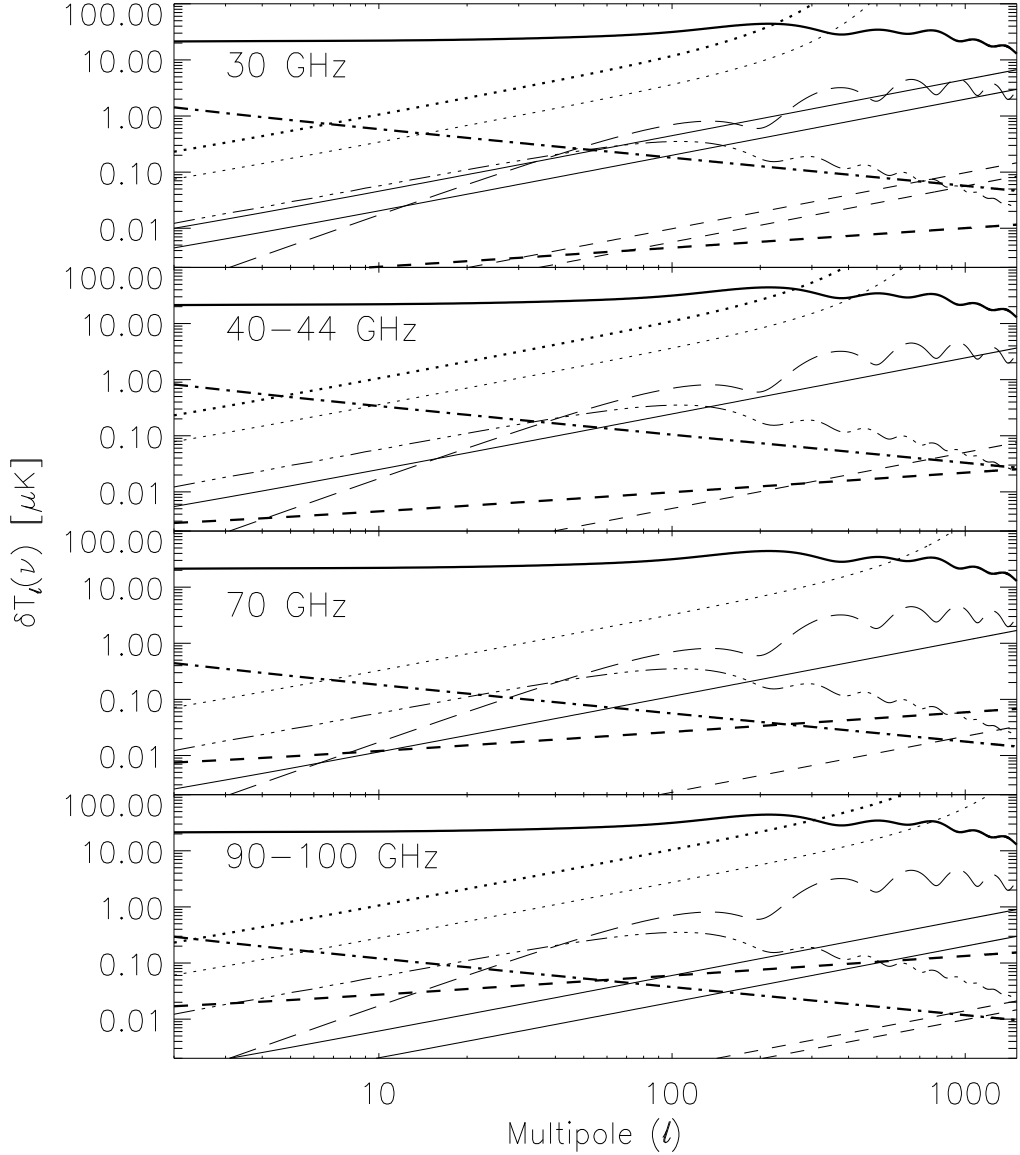,height=14cm,width=14cm} }
\vskip1truecm
\caption{\label{fig2} Power spectra of CMB brightness temperature 
 (thick solid line) and polarization 
fluctuations ($E$-mode: long dashes; $B$-mode: three dots/dash) compared 
with foreground polarization fluctuations at Planck/LFI frequencies. 
Following Tegmark \& Efstathiou (1996) we plot the quantity 
$\delta T_\ell(\nu) = [\ell(2\ell+1)C_\ell(\nu)/4\pi]^{1/2}$. 
The thick dot-dashed and dashed lines correspond to synchrotron and 
interstellar dust emissions, respectively. The thin solid and dashed lines 
show our estimated contributions from extragalactic radio and far-IR 
sources, respectively, based on the model by Toffolatti et al. (1998), as 
updated by De Zotti \& Toffolatti (1999). 
The thin dotted lines represent the expected 
noise spectrum per resolution element, averaged over the sky, of Planck/LFI. 
The heavier dots show the average MAP noise spectrum per resolution element  
at 30, 40 and 90 GHz.}

\end{figure}

\begin{figure}
\vskip-1.3truecm
\centerline{\psfig{figure=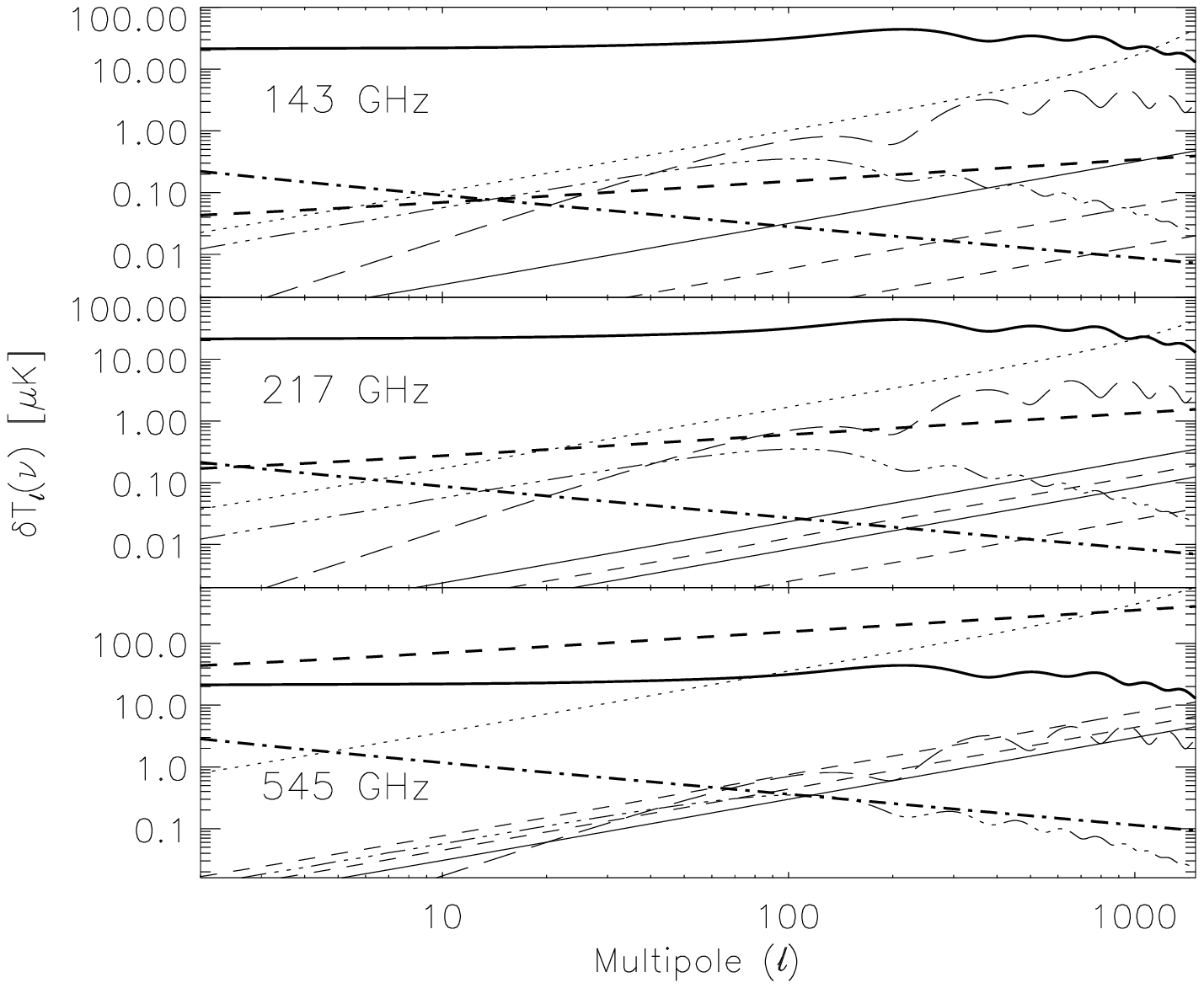,height=12cm,width=14cm} }
\vskip-1.7truecm
\caption{\label{fig3} Power spectra of polarized components at Planck/HFI 
frequencies with sensitivity to polarization. The lines have the same 
meaning as in Fig.~2. We show estimates of polarization fluctuations due to 
dusty galaxies (thin dashed lines) based both on model E by Guiderdoni 
et al. (1998; upper line) and on the model by Toffolatti et al. (1998).}   
\end{figure}

In Figures 2 and 3 the power spectrum of foreground polarization 
fluctuations for all the relevant Planck and MAP channels is 
compared with the power spectrum of CMB anisotropies and of CMB
polarized components. As for the latter, we have plotted the power spectra
of the combinations of Stokes parameters defined by Seljak [1997; his eqs. 
(24) and (25)] and called $E$ and $B$. The estimate of $E$-mode polarization 
fluctuations refer to a standard CDM model 
(scale-invariant scalar fluctuations in a $\Omega=1$, $\Lambda=0$ universe 
with $H_0=50\,\hbox{km}\,\hbox{s}^{-1}\,\hbox{Mpc}^{-1}$ and a baryon 
density $\Omega_b=0.05$). The quantity $B$ vanishes for polarization 
induced by primordial scalar perturbations, and therefore provides a 
unique signature of tensor perturbations (Seljak 1997). 
The $B$-mode power spectrum shown in Figs. 2 and 3 refer to a tilted CDM 
model with power law indices $n_s=0.9$ and $n_t=n_s-1=-0.1$ for scalar 
and tensor perturbations, respectively; the other 
cosmological parameters keep the same values adopted for the standard CDM 
model. Calculations of CMB power spectra 
have been carried out using the CMBFAST package by Seljak \& 
Zaldarriaga (1996).

The thick dashed lines show the $E$-mode of dust polarized power spectrum 
derived by Prunet et al. (1998), scaled to the central frequencies of 
Planck polarized channels using the dust emission spectrum adopted by 
these Authors (emission 
$\propto \nu^2B_\nu(17.5\,\rm{K})$, where $B_\nu(T)$ is the Planck function 
at frequency $\nu$ and temperature $T$). 

The $B$-mode dust 
power spectrum turns out to be close to the $E$-mode one [cf. eqs (11) and 
(12) of Prunet et al. (1998)]. In fact, Seljak (1997) argued that most 
foregrounds should contribute on average the same amount to both modes. 
Therefore, for all foregrounds we consider a single polarized 
power spectrum, assumed to be representative of both modes.

Following Bouchet et al. (1999), we assume that, 
at all frequencies relevant for Planck and MAP, 
the synchrotron polarized emission   
is perfectly correlated with the total synchrotron emission 
(for which a power spectrum $C_T = 4.5 \ell^{-3}\,(\mu\rm{K})^2$ at 100 GHz was 
adopted) and the polarization degree is 44\%. An antenna temperature 
spectral index of 3 ($T_{\rm A, syn} \propto \nu^{-3}$)
has been used to extrapolate the power spectrum (in terms of brightness 
temperature) to the 
other Planck/MAP frequencies (see De Zotti et al. 1999 for references). 

The polarized angular power spectra of extragalactic radiosources 
(thin solid lines) is estimated exploiting the model of   
Toffolatti et al. (1998), as updated by De Zotti \& Toffolatti (1999).
We have assumed that the mean polarization degree of BL Lacs in the 
{\it Stickel north} sample applies to all  
radio galaxies contributing to fluctuations in Planck's channels; based 
on the results shown in Fig. 1, we have adopted a polarization degree 
of 5\% for $\nu \leq 143\,$GHz, of 6\% at 217 GHz and of 10\% at 545 GHz.

As for dusty galaxies (thin dashed lines), we have adopted a polarization 
degree of 2\% at all frequencies and the power spectra of temperature 
fluctuations derived by Toffolatti et al. (1998) and, for HFI channels, 
also by Guiderdoni et al. (1998; their model E).

A flux cut at 1 Jy was adopted for all channels (i.e. sources brighter than 
1 Jy were removed); at 30, 100 and 217 GHz we show also the angular power 
spectrum derived adopting a flux cut of 100 mJy for radiosources 
(lower thin solid lines).

In Table~5 we report the values of $C_\ell$ 
for temperature fluctuations due to a Poisson distribution of 
extragalactic radio sources and dusty galaxies for the same cases shown 
in Figs. 2 and 3. The corresponding values for polarization fluctuations 
follow immediately multiplying by $\Pi^2$. Note that 
a Poisson distribution generates a simple white noise spectrum with the 
same power on all multipoles (Tegmark \& Efstathiou 1996). The values of 
$C_\ell$ are given in terms of brightness temperature 
fluctuations and expressed in $\mu$K$^2$.

\begin{table}
\centering
\caption{Values of $C_\ell$ ($\mu$K$^2$)$^1$ for temperature fluctuations 
for the cases shown in Figs. 2 and 3}
\begin{tabular}{lcccccccc}
 & &  \\ \hline \hline
$\nu$ (GHz)   &     30  &     44  &    70  &    100  &   143  &   217  
&   545  \\
$S_l$ = 1 Jy                           & & & & & & & \\
Radiosources \hspace{-10pt}  &\hspace{-10pt} $5.0(-2)$ & \hspace{-10pt}$1.5(-2)$ 
& 
\hspace{-10pt} $3.2(-3)$ & \hspace{-10pt}$0.9(-3)$ & \hspace{-10pt}$2.5(-4)$ & 
\hspace{-10pt} $9.5(-5)$ &\hspace{-10pt} $5.6(-3)$ \\
(Toffolatti)                 & & & & & & & \\
Far-IR sources &\hspace{-10pt} $1.5(-4)$ &\hspace{-10pt} $4.0(-5)$ & 
\hspace{-10pt} $7.6(-6)$ & $3.1(-6)$ &\hspace{-10pt} $2.8(-6)$ & 
\hspace{-10pt} $1.0(-5)$ &\hspace{-10pt} $3.0(-1)$  \\ 
(Toffolatti)                & & & & & & & \\
Far-IR sources &  & &  &    &\hspace{-10pt} $5.4(-5)$ & \hspace{-10pt}
$2.5(-4)$ &\hspace{-10pt} $8.8(-1)$ \\
(Guiderdoni)       & & & & & & & \\
$S_l$ = 100 mJy  \\
Radiosources &\hspace{-10pt} $1.0(-2)$ &  &  &\hspace{-10pt} $1.0(-4)$ &  &
\hspace{-10pt} $1.2(-5)$ & 
 \\ 
(Toffolatti)                & & & & & & & \\
Far-IR sources &\hspace{-10pt} $5.0(-5)$ &  &  &\hspace{-10pt} $1.5(-6)$ &  & 
\hspace{-10pt} $6.0(-6)$  &   \\
(Toffolatti)                 & & & & & & & \\
Far-IR sources &  &  &  &     &  &\hspace{-10pt} $2.0(-4)$  &  \\
(Guiderdoni)      & & & & & &  \\
\hline \hline
\multicolumn{8}{l}{{}$^1$ In parenthesis are the powers of 10 
(i.e. $5.0(-2)=5.0\,10^{-2}$)}
\end{tabular} 
\end{table} 

As expected, polarization fluctuations due to extragalactic sources are 
particularly relevant at small angular scales. For multipoles $\ell \gsim 
300$, they in fact dominate foreground contributions at 
$\nu \lsim 100\,$GHz. On these scales, at the lowest Planck frequency  
(30 GHz) their amplitude is, according to our estimate, close to that 
of CMB polarization fluctuations induced by scalar perturbations.
In the ``cosmological window'' ($70\lsim \nu \lsim 200\,$GHz), however, 
extragalactic sources are not seriously detrimental to 
measurements of CMB polarization fluctuations.  

The thin dotted lines in Figs. 2 and 3 show the expected power spectra 
of instrumental noise, for polarization measurements,
averaged over the sky, for Planck's LFI and HFI, respectively. 
Following Tegmark \& Efstathiou (1996) we describe the noise power 
spectrum as 
$C_{\ell,{\rm noise}} = \sigma^2 {\rm FWHM}^2 \exp(\ell^2 \sigma_b^2)$ 
where FWHM in expressed in radians, $\sigma_b={\rm FWHM}/2\sqrt{2\ln(2)}$
and $\sigma$ is the rms noise for a square pixel with side FWHM.

For LFI 
we have adopted the sensitivities for brightness temperature measurements 
given by Mandolesi et al. (1998) multiplied by a factor of 2 (Mandolesi, 
private communication). Sensitivities of HFI channels for polarization 
measurements are given by Puget et al. (1998). 
There is a slight difference in the 
mission duration adopted by the two groups to derive their mean sensitivity 
estimates: 12 months for LFI, 14 months for HFI. 

The heavier dots in Fig. 2 
show the expected mean instrumental noise per  resolution element 
for MAP's polarization measurements at 30, 40 and 90 GHz 
obtained from sensitivities for brightness temperature measurements (see the 
MAP Web page) multiplied by a factor $\sqrt 2$ (G. Hinshaw, private 
communication).

As shown by Figs. 2 and 3, the major hurdle in extracting the CMB polarization 
signal is instrumental noise. However, a simple argument shows that, at least 
for a limited band-power range, sufficient sensitivity can be reached. 
We have investigated, in particular, the potential of LFI in 
this respect. The expected sensitivities to polarization per resolution 
element, averaged over the sky, for a 12 months mission, are 6, 10, 14 and 
17$\,\mu$K at 30, 44, 70, and 100 GHz, respectively; the angular resolutions 
(FWHM) are 33', 23', 14', and 10', respectively (Mandolesi et al. 1998).
Simulations done by C. Burigana indicate sensitivities 7 times better 
over areas of about 25 square degrees around each of the ecliptic poles. 
Within these areas, by rebinning the maps at 44, 70 and 100 GHz (we leave aside 
the 30 GHz map which is the most contaminated by polarized foregrounds) 
to a $\simeq 20'$ resolution  
and combining them, a sensitivity to polarization of about $1\,\mu$K can 
be achieved, allowing to image the CMB polarization induced 
by scalar perturbations predicted by the standard CDM model. 
Of course, a lower sensitivity is enough to determine the first moments 
of the distribution of polarization fluctuations.

\section{Conclusions}

We have shown that the polarization fluctuations due to a Poisson 
distribution of point sources with uniform evolutionary properties and 
constant polarization degree are simply equal to intensity fluctuations 
times the average polarization degree. 

The information on the polarization degree of the classes of extragalactic 
sources expected to dominate in the frequency ranges relevant for the MAP 
and Planck missions is scanty. We have taken a conservative approach, 
assuming that all radio sources are as polarized as BL Lac objects and 
that the polarization degree of dusty galaxies is similar to that 
of dense clouds in our own Galaxy and of M$\,$82, a galaxy showing 
a remarkably ordered magnetic field. 

\begin{sloppypar}
We find that, on small scales (multipoles $\ell \gsim 300$), polarization 
fluctuations due to radio sources may indeed dominate foreground 
contributions at $\nu \lsim 100\,$GHz. However, in the ``cosmological 
window'' ($70\lsim \nu \lsim 200\,$GHz), extragalactic sources are not 
a threat for measurements of CMB polarization fluctuations.
\end{sloppypar}

We have also argued that Planck/LFI can reach, in regions around the 
Galactic polar caps, polarization sensitivities of $\simeq 1\,\mu$K, 
allowing to map CMB polarization fluctuations on scales $\sim 20'$. A 
detailed analysis of Planck and MAP capabilities for CMB polarization 
measurements has been carried out by Bouchet et al. (1999).

\begin{ack}
We thank J. Greaves, SCUBA Polarimeter Project 
Scientist, and U. Klaas, from the 
ISOPHOT Data Centre at MPIA Heidelberg, for useful information 
on polarization measurements of external galaxies with SCUBA and ISO, 
respectively. We also thank G. Hinshaw and N. Mandolesi for information on MAP 
and Planck/LFI polarization sensitivities, respectively. T
his work has been done in the 
framework of the Planck-LFI Consortium activities; it has been   
supported in part by ASI, CNR and MURST.
\end{ack}

\end{document}